\documentclass[twocolumn]{article}

\usepackage{amssymb,amsmath,latexsym,amsthm,amsfonts}
\usepackage{graphicx}
\usepackage{mathrsfs}
\usepackage{natbib}

\newenvironment{theacknowledgments}
     {\section*{Acknowledgements}}
     {\par}

\bibpunct{(}{)}{,}{a}{}{,}

\title{{\bf On the relevance of the Bayesian approach to Statistics}}

\author{{\sc Christian P.~Robert}\thanks{C.P. Robert is Professor of Statistics at
Universit\'e Paris-Dauphine, CEREMADE, Paris, France, and Head of the Statistics Lab at
CREST, Paris  Email: \texttt{xian@ceremade.dauphine.fr} Webpage: \texttt{xianblog.wordpress.com}}
}

\begin{document}

\twocolumn 
\maketitle

\begin{abstract}
In this essay, I argue about the relevance and the ultimate unity of the Bayesian approach in a neutral and agnostic manner.
My main theme is that Bayesian data analysis is an effective tool for handling complex models, as proven by the increasing 
proportion of Bayesian studies in the applied sciences. I thus disregard the philosophical debates on the meaning of probability and
on the random nature of parameters as things of the past that ultimately do a disservice to the approach and are irrelevant
to most bystanders.

\end{abstract}

\section{Introduction}
Bayesian data analysis can be defined as {\em a method for summarising uncertainty and making 
estimates and predictions using probability statements conditional on observed data and an assumed model} 
\citep{gelman:2008}. In this essay, I aim to explain why I believe (with many others) that Bayesian data analysis
is valuable and useful in statistics, econometrics, and biostatistics, among other fields. My defence
of the theme is based on presenting a user's perspective and arguing in favour of the ultimate practicality of the Bayesian
toolbox, whilst refraining from more elaborate philosophical and epistemological arguments on the nature of Science.

I do agree with Russell Davidson that the shrill tone of some---mostly past---defences of the Bayesian paradigm are doing it a
disservice by transferring the debate to religious and therefore irrational grounds.\footnote{The barb of Russell Davidson,
also found in~\cite{senn:2008},
{\em  Bayesians are of course their own worst enemies. They make non-Bayesians accuse
them of religious fervour, and an unwillingness to see another point of view}, is not completely unfounded.}
My personal stance on the Bayesian choice is on the contrary grounded
in realism. The Bayesian perspective provides me with a complete toolbox that allows me 
to conduct inference in an arbitrary setting at a minimal cost in terms of constructing statistical procedures. 
In addition, it provides sufficient theoretical safety
rails to ensure coherence in my decision-making and convergence properties for my procedures. I also agree with Andrew Gelman's
(\citeyear{gelman:2008}) reservation that {\em a consequence 
of Bayesian statistics being given a proper name is that it encourages too much historical deference
from people who think that the bibles of Jeffreys, de Finetti, or Jaynes have all the answers}.
The formalisation of Bayesian statistics by those pioneers has greatly contributed towards more efficiency
in the design of Bayesian procedures \citep{robert:chopin:rousseau:2009} and therefore to their current popularity. 
However, naming a technique after particular scientists, even when as prestigious as those above, is a rhetorical trick to bring 
more authority to an approach. To keep the tone of this essay as
clear as possible, I will nonetheless use the recent \citep{fienberg:2006} adjective of ``Bayesian" in the following
but I will mostly refrain from giving a name to alternatives, the usual adjective of ``frequentists" seeming now out-dated and 
overly restrictive. The range of non-Bayesian statistical techniques indeed extends much further than looking at average properties.

As already done in the above, throughout the text I will be making (an admittedly selective) use of recent quotes 
that defend or criticise the Bayesian approach. Most of 
them emanate from a debate run by {\em Bayesian Analysis} following the tongue-in-cheek critique of \cite{gelman:2008}.
I will present here elements to support Gelman's (\citeyear{gelman:2008}) conclusion that, {\em given the advances in practical 
Bayesian methods in the past two decades, anti-Bayesianism is no longer a serious option}. My view is that 
denying the relevance of Bayesian analysis on the sole ground that it is Bayesian does not follow from a rational stance.

\section{Bayesian models}
Let me first stress that the Bayesian approach to non-parametrics is alive and well, as shown for instance by the recent advances in Dirichlet
models \citep{teh:jordan:beal:blei:2006} and Bayesian asymptotics \citep{ghosal:vaart:2006} (see also 
\citealp{hjort:holmes:mueller:2009}). Bayesian non-parametrics
can now manage density and functional estimation with the same degree of complexity
with which a normal mean is estimated by a Bayesian analysis based on a conjugate prior \citep{denison:holmes:mallick:smith:2002}. 
As regards Russell Davidson's first question related to the Bahadur-Savage impossibility theorem, 
I do not understand the statistical point of the test in his Section 3 and I therefore have
no answer. (His Theorem 1 reminds me very much of a result of the late Costas Goutis, reported in my book, \citealp{robert:2001},
Table 3.2.3, about the range of Bayes estimators.) On the other hand, the issue
raised by Russell Davidson in Section 7 about incorporating smoothness in the prior does not seem to be particularly problematic,
once smoothness is defined in terms of a particular class of functions.

I will only consider here parametric settings, mostly for simplicity and space reasons. (And also for the fact that the 
priors found in non-parametric settings seem to be much more acceptable as working tools by non-Bayesians.) 
The common ground for both parametric and non-parametric settings is nonetheless
that a model provides a likelihood. I simply do not believe meaningful inference is possible without 
this likelihood function.\footnote{Of course, this statement goes against a large portion of the current
practice that contends that first moments are sufficient descriptions of the real world. But I do prefer
the facilities provided by a full if wrong model to the adhocqueries required by a minimalist modelling.
In particular, replying to Russell Davidson's question in Section 4,
I do not think there is a Bayesian approach to GMM's unless one is ready to use a pseudo-likelihood
that encompasses the specified moments.}

Given that all models are approximations of the real world, the choice of a parametric model obviously is wide-open to criticism.
As stated by \cite{gelman:2008},
{\em Bayesians promote the idea that a multiplicity of parameters can be handled via hierarchical, typically exchangeable, models, but
it seems implausible that this could really work automatically [instead of] giving reasonable answers using minimal assumptions}.
This is, however, a type of criticism that goes beyond Bayesian modelling {\em per se} and questions the relevance of
completely built models for drawing inference or running predictions. (Obviously, embracing my ``opponent's" perspective 
that inference is sometimes impossible would immediately close the discussion!) The Bayesian paradigm does not state that the model with
which it operates is the ``truth", no more than it requires that the corresponding prior distribution has a connection
with the ``true" production of parameters (since there may even be no parameter at all). It simply provides an inferential machine that
has strong optimality properties under the right model and that can similarly be evaluated under any other well-defined alternative models. 
In Popper's (\citeyear{popper:1934}) terms, a Bayesian model can be ``falsified" when faced with data from another 
model.\footnote{This is not to imply that the philosophy of \cite{popper:1934} is in agreement with the Bayesian approach, since
\cite{popper:miller:1983} demonstrates the impossibility of coherent statistical inference.} \cite{templeton:2008}
sees the fact that {\em having a high relative {\em [posterior]} probability does not mean that 
a hypothesis is true or supported by the data} as the ultimate drawback of the Bayesian paradigm. 
On the contrary, I see it as a strength, even in Popperian terms, because (a) there is no such thing as 
a ``true" hypothesis and (b) the support brought by the data is always relative to a reference model. 
Besides, the Bayesian approach is such that {\em techniques allow prior beliefs to be tested and discarded 
as appropriate} \citep{gelman:2008}. In other words, {\em Bayesian data analysis has three stages: formulating a model, fitting the model 
to data, and checking the model fit} \citep{gelman:2008}. Hence, there seems to be little reason for not using a parametric model at an
early stage even if it is later dismissed as ``not true enough" (in favour of another model). 

Besides giving the Bayesian paradigm his name,
Thomas Bayes contributed by stating the {\em definition} of a conditional probability
and deriving what is now known as Bayes theorem.\footnote{As stressed by \cite{jaynes:2003},
Bayes' contribution to inference was essentially restricted to a somewhat dubious toy
example of locating the position of a billiard ball. In contrast, Laplace and others had a much wider range of examples,
with more realistic applications. Jeffreys, de Finetti and Jaynes set the bases into firmer mathematical and methodological ground, 
while Wald and Stein established the fundamental optimality properties validating Bayes procedures.}
Nonetheless, if surprisingly, there still exists a debate about the very nature of Bayes 
theorem. Russell Davidson points out that it is difficult to express it {\em
in the formalism that is used in financial economics/econometrics}. 
Another illustration is given by \cite{templeton:2008}. He argues that conditioning upon the observation $x\sim f(x|\theta)$
is plainly invalid: {\em The impact of treating $x$ as a fixed constant
is to increase statistical power as an artefact} and {\em ignoring the sampling error of $x$ undermines
the statistical validity of all inferences made by the method}. As validated by standard measure theory \citep{billingsley:1986},
the posterior distribution
\[
\pi(\theta|x) = {f(x|\theta) \pi(\theta) \over \int f(x|\theta)
     \pi(\theta) \,\text{d}\theta}
\]
does include the sampling (or {\em error}) distribution while conditioning on the data $x$. This 
approach furthermore is the only coherent way
to give a meaning to statements like $\mathbb{P}(\theta>0|x)$, i.e.~to properly construct confidence and prediction statements, while
conditioning on the data at hand.\footnote{This point relates to Russell Davidson's questions about the bootstrap. While I appreciate
very much the strength of bootstrapping techniques and find them a natural entry to Statistics for my third year students, I have
trouble reconciliating the bootstrap and Bayesian statistics. Indeed, the bootstrap is fundamentally a plug-in method, especially in
its parametric version, which therefore omits to properly take into account the variability of the plugged-in parameter estimates.}

\cite{gelman:2008} reports that {\em Bayesian methods are presented as an automatic inference engine, 
and this raises suspicion in anyone with applied experience}. It is true that $\pi(\theta|x)$ is the core of Bayesian inference.
It can legitimately be viewed as the ``ultimate inference engine" via which all decisions (in a decision-theoretic framework) 
based on the data can be automatically derived. There is no fundamental difficulty in this automated 
derivation.\footnote{That it is an automatic engine is an argument rarely advanced by critics of the Bayesian approach, 
who on the contrary uniformly point out its subjective features. See Section \ref{sec:witch}.} 
Once optimality criteria are explicitly stated via the utility function associated with the decision, searching for the optimal decision
reduces to solving a well-posed optimisation problem.\footnote{Gelman (\citeyear{gelman:2008}) stresses that  
{\em loss functions [are] not relevant to statistical inference} and he does not
{\em see any role for squared error loss, minimax, or the rest of what is sometimes called statistical decision theory}. 
Following the arguments advanced in \cite{robert:2001},
but also in \cite{berger:1985} and \cite{bernardo:smith:1994}, I cannot but strongly disagree with this perspective.
Decision theory is a strong motivation for using Bayesian procedures, especially in economics and econometrics where rationality
is customarily associated with maximising utility functions.}
Furthermore, the {\em inference {\em [step]} gets most of the attention, but the Bayesian procedure as a 
whole is not automatic} \citep{gelman:2008}. In addition, using a probability distribution on the parameter space and Bayes theorem allows for a
coherent update of the information available on $\theta$ in the sense that the current posterior distribution becomes the prior distribution 
before gathering more data.

\section{On prior selection}\label{sec:witch}
The recurrent criticism of the Bayesian perspective is that the whole inferential approach is ultimately dependent upon the choice of the prior
distribution, as clearly shown by the definition of the posterior distribution above. There is no possible debate about this fact, 
either from a mathematical or methodological perspective. It is also straightforward to come up with examples where the choice of the
prior leads to absurd decisions. 

There is no easy answer to this criticism, but this acknowledgement must not be taken as conceding 
defeat in the debate! If the prior had no impact on the inference, data would be similarly useless, since the 
update would not matter. Therefore, I see this dependence as a plus of the Bayesian approach. It allows one to include 
an infinite range of prior opinions and items of information, while progressively concentrating on neighbourhoods of the 
``true" value of the parameter---in settings where the data is generated from the assumed model. 
In the literature, this point about the advantages of
incorporating prior information is rather universally accepted. The criticisms instead focus on the opposite situation
where the prior information is poor or inexistent, denying {\em non-informative} (or {\em ignorance})
priors their label, i.e.~the representation of a state of complete ignorance.

Maybe surprisingly (and maybe not!), I completely agree with this criticism in that any choice of prior distribution 
corresponds to some informational input about the parameter. The ultimate argument is that,
were there such a thing as {\em the} non-informative prior, it would {\em be expected to 
represent total ignorance about the problem} \citep{kass:wasserman:1996}. Thus, being moderately
unfair (!), this object should be such an information black
hole as to cancel the effect of any amount of information and should thus remain the same even after observing the data! Therefore, when 
\cite{jeffreys:1939} states that {\em if the parameter may have any value from $-\infty$ to $+\infty$, its prior probability should be taken 
as uniformly distributed}, he is making a choice of a particular structure of the model that impacts on his future inference, in addition
to using the term {\em uniform} in an implicitly generalised manner because the parameter space is then unbounded
\citep{robert:chopin:rousseau:2009}. Instead, as stated by \cite{gelman:2008},
{\em there is no good objective principle for choosing a noninformative prior (even if that concept were mathematically defined, which it is 
not)}. The notions of {\em objective} and of {\em non-informative} are indeed not well-defined mathematical concepts and they carry 
an irrational undertone that fails to lend legitimacy to the associated priors. Some mathematical criteria do lead to some 
competing families of {\em reference} priors like the left Haar measures mentioned by Russell Davidson or matching priors (see 
\citealp{robert:2001}, Chapters 3 and 8). The ultimate attempt at producing a meaningful rationale for building non-informative
priors is, in my opinion,
Bernardo's (\citeyear{bernardo:1979}) definition through the information theoretical device of Kullback divergence (see also 
\citealp{berger:bernardo:1992}). Quite obviously, this is not the only possible approach. Among other things, it depends on a choice of
information measure, does not always lead to a solution and requires an ordering of the model parameters that involves some prior information
(or some subjective choice). However, as long as we do not {\em think of those reference priors as representing ignorance}  \citep{lindley:1973},
they can indeed be {\em taken as reference priors, upon which everyone could fall back when the prior information is missing}
\citep{kass:wasserman:1996}. 

Apart from the conceptual confusion about non-informative priors that plagued most of the 19th and mid 20th century debate about
the nature of Bayesian inference, the issue of improper priors often serves as a further criticism. Indeed, non-informative priors
often are measurable functions $\pi(\theta)$ with infinite mass,
$$
\int_{\Theta} \pi(\theta)\,\text{d}\theta = +\infty\,,
$$
which deprives them of a probabilistic interpretation. This criticism can be most easily 
rebutted for a wide variety of reasons. The first reason is topological coherence: limits of
Bayesian procedures often partake of their optimality properties \citep{wald:1950} and
should therefore be included in the range of possible procedures. Another one is robustness: a measure
with an infinite mass is much more robust than a true probability distribution with a large
variance. Provided
$$
\int_{\Theta} f(x|\theta) \pi(\theta) \,\text{d}\theta < \infty\,,
$$
the quantity
$$
\pi(\theta|x) = {f(x|\theta) \pi(\theta) \over
           \int_{\Theta} f(x|\theta) \pi(\theta) \,\text{d}\theta}
$$
is as well-defined as a probability density as a regular posterior 
distribution \citep{hartigan:1983,berger:1985,robert:2001}.

\section{Testing versus model comparison}
The inferential problems of Bayesian model selection and of Bayesian testing are clearly those for which the most vigorous criticisms can
be found in the literature. An illustration is provided by \cite{senn:2008} who states that {\em the Jeffreys-subjective synthesis betrays 
a much more dangerous confusion than the Neyman-Pearson-Fisher synthesis as regards hypothesis tests}. I find this suspicion rather
intriguing given that the Bayesian approach is the only one giving a proper meaning to the probability of a null hypothesis,
$\mathbb{P}(H_0|x)$. Alternative methodologies are able,  at best, to specify a probability value on the {\em sampling} space, i.e.~on
the ``wrong" space since the only variation is on the parameter space once the observation is obtained.

\cite{senn:2008} further advances that {\em what is almost never used, however, is the Jeffreys significance test}. I recall
here that the most standard Bayesian approach to testing and model choice relies on the Bayes factor \citep{kass:raftery:1995},
which, for hypotheses written as $H_0:\,\theta\in\Theta_0$ and as $H_1:\,\theta\in\Theta_1$, is defined as \citep{jeffreys:1939,jaynes:2003}
$$
B_{01} = \displaystyle{ \frac{\pi(\Theta_0|x)}{ \pi(\Theta_1|x)} \bigg/
      \frac{\pi(\Theta_0) }{ \pi(\Theta_1)} }
  = \frac{ \displaystyle{\int_{\Theta_0} f(x|\theta) \pi_0(\theta) \text{d}\theta} }{
       \displaystyle{\int_{\Theta_1} f(x|\theta) \pi_1(\theta) \text{d}\theta} }\,.
$$
This monotonic transform of the posterior probability of $H_0$ eliminates the influence of the prior weight $\pi(\Theta_0)$
and has a similar interpretation to the classical likelihood ratio. However, it does not suffer from the over-fitting
difficulties of the latter, in that it includes a natural penalisation factor for richer models. This is shown by the connection
with the BIC (Bayesian information criterion), intuited by \cite{jeffreys:1939}: {\em variation is random until the 
contrary is shown; and new parameters in laws, when they are suggested, must be tested one at
a time, unless there is specific reason to the contrary}. Although I strongly dislike using the term 
because of its undeserved weight of academic authority, the Bayes factor acts as a natural {\em Ockham's razor.}

A criticism of the use of Bayes factors (e.g., \citealp{templeton:2008})
is that the quantity is not scaled in probability terms. On the contrary, I maintain
it is naturally scaled against one and can, moreover, be readily transformed into posterior probabilities 
when the prior probabilities of the hypotheses
are specified. (It is furthermore a natural factor in a decision-theoretic framework, see \citealp{robert:2001}.) 
Another criticism is rarely voiced outside the Bayesian community, namely that the use of improper priors is mostly
prohibited in this setting, for lack of proper normalising constants. Solutions have been proposed, akin to cross-validation techniques
in the classical domain \citep{berger:pericchi:1996,berger:pericchi:varshavsky:1998}, but they are somehow too ad-hoc to convince the
entire community (and obviously beyond). 

If we consider the special case of point null hypotheses---which is not so limited in scope since it includes all variable
selection setups---, there is a difficulty with using a standard prior in this environment. As put by \cite{jeffreys:1939},
when {\em considering whether a location parameter $\alpha$ is $0$ {\em [when]} the prior is uniform, we should have to take 
$\pi(\alpha)=0$ and $B_{10}$ would always be infinite}. This is a case when the inferential question implies a
modification of the prior, justified by the information contained in the question. Avoiding the whole issue is a clear-cut solution, as 
with Gelman (\citeyear{gelman:2008}) having {\em no patience for statistical methods that assign positive probability to point 
hypotheses of the $\theta = 0$ type that can never actually be true}. Considering the null and the alternative hypotheses as
defining two different models is another solution that allows for a Bayes factor representation. 

A major criticism directed at the Bayesian approach to testing is that it is not interpretable on the same scale as
the Neyman-Pearson-Fisher solution, namely in terms of Type I error probability and test power. In other words, 
{\em frequentist methods have coverage guarantees; Bayesian methods don't; 95 percent frequentist intervals will live up to their 
advertised coverage claims} \citep{wasserman:2008}. A natural thing to do is then to question the appeal of such frequentist
properties when considering a single dataset. That is, in Jeffreys' (1939) famous words, {\em a hypothesis that may be true
may be rejected because it had not predicted observable results that have not occurred}. 
From a decision-theoretic perspective---to which the
frequentist properties should relate---, a classical Neyman-Pearson-Fisher procedure is never evaluated in terms of the consequences
of rejecting the null hypothesis, even though the rejection must imply a subsequent action towards the choice of an alternative model.
(From a narrower decision-theoretic perspective, note also that
$p$-values may be inadmissible estimators, \citealp{hwang:casella:robert:wells;farrel:1992}.)
Therefore, arguing that high posteriors probabilities do not imply that a hypothesis is true as in
\cite{templeton:2008} and that the Bayesian approach is relative in that it {\em posits two or more alternative hypotheses and 
tests their relative fits to some observed statistics} \citep{templeton:2008}, is missing the main purpose of 
Bayesian tests. Bayesian procedures do not aim at validating or invalidating a golden model {\em per se} but rather 
lead to the choice of a working model that allows for acceptable predictive 
properties.\footnote{It is worth repeating the earlier assertion that all models are false and that finding that a hypothesis 
is ``true" is not within our reach, if at all meaningful!}

Another criticism covers the lack of asymmetry of the Bayes factor, since it satisfies the
equality $B_{10}=1/B_{01}$. For model choice, i.e.~when several models are under comparison for the same observation
$$
\mathfrak{ M}_i : x \sim f_i(x|\theta_i)\,, \qquad i \in \mathfrak{I}\,,
$$
where $\mathfrak{I}$ can be finite or infinite, this symmetry seems to me to be a fundamentally sound property. Nevertheless, 
Templeton (2008) bemoans that {\em there is no null hypothesis, which complicates the computation of sampling error, since there is no single 
statistical model under which to evaluate sampling}. This should be construed as a clear limitation of the Neyman-Pearson-Fisher
paradigm, since the latter imposes asymmetry and (Type I) error control under a single (null) model. 
However, this is not the perspective of Templeton (2008)
who concludes with the impossibility of the posterior probability of a model,
$$
\pi({\mathfrak M}_i|x) = \dfrac{ \displaystyle{p_i \int_{\Theta_i} f_i(x|\theta_i)
         \pi_i(\theta_i) \text{d}\theta_i}
}{
\displaystyle{\sum_j p_j \int_{\Theta_j} f_j(x|\theta_j) \pi_j(\theta_j)
      \text{d}\theta_j}  }
$$
due to the impression that {\em the numerators are not co-measurable across hypotheses, and the denominators
are sums of non-co-measurable entities. Hence, the ``posterior probabilities" that emerge are not 
co-measurable.  This means that it is mathematically impossible for them to be probabilities.}
Given that all terms are marginal likelihoods for the same observation, it seems difficult to argue against
their co-measurability. Contrary to classical plug-in likelihoods, marginal likelihoods do allow for a comparison on the same scale.
Similarly, the belief that {\em complicating dimensionality of test statistics is the fact that the models
are often not nested, and one model may contain parameters that do not have analogues in the other models and vice versa}
\citep{templeton:2008} is not well-founded. The Bayes factor is properly defined and applicable to settings where the models are
not embedded (or nested). This is due to the fact that the corresponding quantity of interest for a given model is the marginal likelihood
(or evidence),  which integrates over spaces and complexity and which can be interpreted at face value since it is calibrated across models.

A last point of contention about Bayesian testing is the apparent absence of clearly defined directions when conducting a standard analysis. 
Figure \ref{fig:from:Core} reproduces an output from \cite{marin:robert:2007}. This computer output illustrates how a default prior 
and Bayes factors can be used in the same spirit as significance levels in a standard regression model, 
each Bayes factor being associated with the test 
of the nullity of the corresponding regression coefficient. This output mimics the standard {\sf R} function {\sf lm} outcome in order to 
show that the level of information provided by the Bayesian analysis goes beyond the classical output. My point here is obviously
{\em not} in showing that we can get similar answers to those of a least square analysis since, else, {\em 
we might as well use the frequentist method} \citep{wasserman:2008}.  It is to demonstrate that reference analyses are available, while
preserving the strength of the Bayesian machinery (like joint confidence regions and multiple tests).

\begin{figure}
{\sffamily

\begin{tabular}{l l l l}
            &Estimate  &BF        &log10(BF)\\
& & & \\
(Intercept)  &9.2714   &26.334  &1.4205 (***) \\
X1          &-0.0037   &7.0839  &0.8502 (**) \\
X2          &-0.0454   &3.6850  &0.5664 (**) \\
X3          &0.0573    &0.4356  &-0.3609 \\
X4          &-1.0905   &2.8314  & 0.4520 (*) \\
X5          & 0.1953   &2.5157  & 0.4007 (*) \\
X6          &-0.3008   &0.3621  &-0.4412 \\
X7          &-0.2002   &0.3627  &-0.4404 \\
X8          & 0.1526   &0.4589  &-0.3383 \\
X9          &-1.0835   &0.9069  &-0.0424 \\
X10         &-0.3651   &0.4132  &-0.3838 \\
\end{tabular}

\medskip
evidence against H0: (****) decisive, (***) strong,
(**) substantial, (*) poor
}

\caption{\label{fig:from:Core} {\sf R} output of a Bayesian regression on a processionary caterpillar
dataset with ten covariates analysed in \cite{marin:robert:2007}.}
\end{figure}

\section{On pervasive computing}
Bayesian analysis has long been derided for providing optimal answers that could not be computed. With the advent of
early Monte Carlo methods, of personal computers, and, more recently, of more powerful Monte Carlo methods \citep{hitchcock:2003}, the
pendulum appears to have switched to the other extreme. Nowadays, {\em Bayesian methods seem to quickly move to elaborate computation}
\citep{gelman:2008}. This feature does not make Bayesian methods less suspicious in the mind of critics, for different reasons:
{\em a simulation method of inference hides unrealistic 
assumptions} \citep{templeton:2008}. I won't launch here into a defence of simulation techniques that have done so much to
promote Bayesian analysis in the past decades, referring to \cite{chen:shao:ibrahim:2000,
robert:casella:2004,marin:robert:2007} for detailed arguments and to \cite{robert:marin:2009,robert:wraith:2009} for specific
coverages of the computational advances related to Bayesian model choice. Simulation methods can certainly be misused---as 
any methodology can be---. However, while {\em Bayesian simulation {\em [may seem]} stuck in an infinite regress of inferential uncertainty} 
\citep{gelman:2008}, there exist enough convergence assessment techniques \citep{robert:casella:2010} to ensure a reasonable 
degree of confidence 
in the accuracy of the approximation provided by those simulation methods. Thus, as rightly stressed by \cite{bernardo:2008}, 
{\em the discussion of computational issues should not be allowed to obscure the need for further analysis of inferential 
questions}.\footnote{The confusion of \cite{templeton:2008} is of this nature, namely his criticisms bear in fact on the 
generic principles of Bayesian inference and in particular testing
while he aims at criticising a specific simulation methodology called ABC and described below. See \cite{beaumont:etal:2010}
for a discussion of this confusion.}

In Section 6, Russell Davidson asks about the reliability of Markov chain Monte Carlo (MCMC) methods and about recent developments 
in this field. The answer is more complex than time and space allow in this essay, so my first reply is to refer him to 
\citep{robert:casella:2004,robert:casella:2009} for booklength entries. A second response is that, despite their specific label,
MCMC methods do not differ in essence from other Monte Carlo methods. When using an importance sampler or an harmonic mean
estimator (see \citealp{marin:robert:2010} for details), the quantities we produce are unbiased, which is not a characteristic
of MCMC outputs. However, they may also be associated with infinite variance, which means that their convergence time is beyond
anyone's patience! The same applies to MCMC samples which are formally associated with the correct stationary distribution but which
may in practice end up with a cosmological number of iterations! \cite{robert:casella:2010} details several tools that help in
checking convergence and stationarity, but those tools are not completely foolproof. Therefore it may happen that the lack of convergence
of a MCMC output remains undetected. Similarly, using a numerical integration software may fail to detect an important region for the
integrand. Those are numerical problems that have little to do with the methodology under scrutiny and can often be detected by using
a multifaceted strategy, mixing together several numerical methods.

Interestingly enough, the most
accurate---in our opinion---approximation technique for Bayes factors is, when applicable, derived from Bayes theorem, 
This is indeed the purpose of Chib's (\citeyear{chib:1995}) rendering:
$$
m(x) = \dfrac{\pi(\theta) f(x|\theta)}{\pi(\theta|x)} \approx \dfrac{\pi(\theta) f(x|\theta)}{\hat\pi(\theta|x)}\,, 
$$
where $\hat\pi(\theta|x)$ is a simulation-based approximation to the posterior density. 
\cite{marin:robert:2008} propose an illustration in the setting of mixtures, while \cite{robert:marin:2009} implement
the method for a probit model, with both examples demonstrating the precision of this approximation,
There have been discussions about the accuracy of this method in multimodal settings \citep{fruhwirth:2004}, but straightforward 
modifications \citep{berkhof:mechelen:gelman:2003,lee:marin:mengersen:robert:2008} overcome such difficulties and make for both 
an easy and a robust computational tool associated with Bayes factors. 

Instead of presenting the whole range of available computational solutions, I want to point out here a single but recent advance in 
Bayesian computing that allows for a further extension of Bayesian data analysis to cases
where any other method of inference is either impossible or seriously inaccurate. This new method is called ABC, standing for
{\em Approximate Bayesian Computation}. It was introduced in genomics by \cite{pritchard:seielstad:perez:feldman:1999} to handle
models, like phylogenic trees, where the likelihood could not be computed in a reasonable time, hence prohibiting the use of standard
simulation tools.  The method is based on a standard accept-reject principle generating $\theta\sim\pi(\theta),x^\prime\sim 
f(x|\theta)$ until $x^\prime=x$ which produces a generation from $\pi(\theta|x)$. Since the
stopping rule is impossible to attain in continuous settings, the approximation in ABC consists in replacing $x=x^\prime$ with a
relaxed condition, $d(x,x^\prime)<\epsilon$, where $d$ is an arbitrary divergence measure and $\epsilon$ is an approximation 
parameter to be calibrated..  Assuming that new ``observations" $x^\prime$ from the likelihood can be easily simulated, this 
method provides controlled approximations $\pi(\theta|d(x,x^\prime)<\epsilon)$
to the posterior distribution. The accuracy of this method can be calibrated against the available computing power and it is 
currently in standard use for genomic applications \citep{cornuet:santos:beaumont:etal:2008} as well as for model choice in graphical models
\citep{grelaud:etal:2009}.\footnote{\citep{grelaud:etal:2009} 
is one illustration of the high popularity of Bayesian techniques in epidemiology,
biostatistics and genomics. I thus disagree with Russell Davidson's impression of the opposite at the end of Section 8!} 

The field of Bayesian computing is therefore very much alive and, while its diversity can be construed as a drawback
by some, I do see the emergence of new computing methods adapted to specific applications as most promising, because it bears 
witness to the growing involvement of new communities of researchers in Bayesian advances.

\section{Conclusion}
Once again, I want to stress that the purpose of this essay is far from trying to preach in favour of my creed, as I do not
see Bayesian data analysis as a philosophical (and even less religious) stance. What drives my Bayesian choice is the essential
practicality of the tools and of the actions I can undertake thanks to that choice, as well as the ability to evaluate, criticise, and
possibly modify, the calibration choices I have made at the beginning of my analysis. There is beauty as well as efficiency in 
transparency and a Bayesian data analysis is ultimately transparent in that it displays all of its components (prior, likelihood, loss
function, simulation technique) for public evaluation. The fact that any of these components can be replaced by an alternative
version explains illustrates the versatility of the method and the appeal it exerts on non-statisticians in need of a data analysis
tool. The other practical side of Bayesian data analysis is that we now see a growing range of complex models where, apart 
from abdicating on some part of the complexity, the only available solution is to use a Bayesian approach. Handling highly 
non-identifiable models, inferring about the graphical structure of a spatial model, running a small area estimation on an 
very dense grid, analysing continuous time data with hidden Markov structures, all of these problems and a myriad of others cannot
be processed but from a Bayesian perspective.


\begin{theacknowledgments}
C.P.~Robert is supported by the ANR-2009-BLAN-0318 ``Big'MC" grant.
He is grateful to the {\em Bayesian Econometrics III workshop} organisers for their invitation to lovely
Rimini and for their support, and to Russell Davidson for such a congenial and open discussion. The editing 
help of Gael Martin over two successive versions greatly improved the presentation of this essay. Comments
from Nicolas Chopin are also gratefully acknowledged.
\end{theacknowledgments}

\small
\bibliographystyle{ims} 

\IfFileExists{\jobname.bbl}{}
 {\typeout{}
  \typeout{******************************************}
  \typeout{** Please run "bibtex \jobname" to obtain}
  \typeout{** the bibliography and then re-run LaTeX}
  \typeout{** twice to fix the references!}
  \typeout{******************************************}
  \typeout{}
 }

\end{document}